**Mind the Gap: Securely modeling cyber risk based on security deviations from a peer group**

Authors: Taylor Reynolds, Sarah Scheffler, Daniel J. Weitzner, Angelina Wu[1]

## Abstract

There are two strategic and longstanding questions about cyber risk that organizations largely have been unable to answer: What is an organization's estimated risk exposure and how does its security compare with peers? Answering both requires industry-wide data on security posture, incidents, and losses that, until recently, have been too sensitive for organizations to share. Now, privacy enhancing technologies (PETs) such as cryptographic computing can enable the secure computation of aggregate cyber risk metrics from a peer group of organizations while leaving sensitive input data undisclosed. As these new aggregate data become available, analysts need ways to integrate them into cyber risk models that can produce more reliable risk assessments and allow comparison to a peer group. This paper proposes a new framework for benchmarking cyber posture against peers and estimating cyber risk within specific economic sectors using the new variables emerging from secure computations. We introduce a new top-line variable called the "Defense Gap Index" representing the weighted security gap between an organization and its peers that can be used to forecast an organization's own security risk based on historical industry data. We apply this approach in a specific sector using data collected from 25 large firms, in partnership with an industry ISAO[2], to build an industry risk model and provide tools back to participants to estimate their own risk exposure and privately compare their security posture with their peers.

---

[1] Authors listed in alphabetical order. Weitzner and Wu were supported, in part, by NSF grant Collaborative Research: DASS: Legally Accountable Cryptographic Computing Systems (LAChS) Award Number: 21315415. Reynolds was supported by MIT's Future of Data Initiative, MIT's FinTech@CSAIL, and MIT's Machine Learning Applications @CSAIL.

[2] The data was collected from 25 large firms in the United States with combined annual revenues of USD 23 billion. Due to the sensitive nature of the results, we are keeping the name of the ISAO undisclosed in this version of the paper.

# Introduction

There are two strategic and longstanding questions about cyber risk that organizations largely have been unable to answer: What is an organization's estimated risk exposure and how does its security compare with peers? Answering both requires industry-wide data on security posture, incidents, and losses that, until recently, have been too sensitive for organizations to share.

Until now, firms have been unable to assess their own cyber risk posture with reference to larger risk patterns. This means that firms have been unable to forecast their own cyber risk because they lack the tools to learn about the frequency and magnitude of attacks in their own sector and the defensive posture of their peer group. Some of this data has been narrowly available to insurers, but even insurance providers and brokers lack data on core data such as actual economic losses (in contrast with insured losses) that are necessary to accurately forecast cyber risk. This lack of data leaves organizations struggling to answer basic questions about the magnitude of their own cyber risk and how they compare with other organizations in their peer group.

This comes at a time when government regulations increasingly require organizations to evaluate and monitor their cyber risk and the effectiveness of security controls. For example, the newly revised FTC Safeguards Rule in the United States requires organizations, now extending beyond just financial services, to conduct security risk assessments that "must be written and must include criteria for evaluating those risks and threats."("FTC Safeguards Rule: What Your Business Needs to Know" 2022). The New York State Department of Financial Services recently issued a rule requiring covered entities to confirm that they have devoted adequate resources to cover expected risk (NYDFS 2023). In Europe, Article 21 of the European Union's Network and Information Security Directive (NIS 2) mandates that organizations have "policies on risk analysis and information system security" as well as "policies and procedures to assess the effectiveness of cybersecurity risk-management measures"(EU 2022).

It is not just governments putting new demands on organizations to produce cyber risk assessments and track the effectiveness of controls. The US National Association of Corporate Directors (NACD) produced a 2023 Director's Handbook on Cyber-Risk Oversight that calls for management to "deliver reports that are benchmarked, so directors can see metrics in context to peer companies or the industry" (NACD 2023). In addition, the NACD says directors should obtain cyber risk assessments and information about cyber-risk exposure in economic terms (NACD 2023).

While there are clear policy requirements for organizations to evaluate the effectiveness of controls and compare themselves to peers, the ability to do so must be called into question without access to the key external data about their peers that they would need to so effectively. This could change though as new cryptographic techniques open access to aggregated cyber security data within a particular industry.

Cryptographic computation tools, a type of privacy enhancing technology or PETS, facilitate new data sources within an industry that can be used to benchmark and model risk. A subset of PETs known as "encrypted data processing tools" or "cryptographic computing" allow aggregated results to be computed from encrypted cyber security posture, incident, and loss data without requiring organizations to disclose the individual inputs. These secure computation approaches are used to develop cybersecurity benchmarks that can be used by individual firms for private comparisons (de Castro et al. 2020).

The introduction of secure computation techniques for data analytics opens access to data sets that were never available before, particularly at the sector level among a group of peers. Organizations can now share sensitive information into a computation without the risk of revealing or disclosing sensitive, proprietary, or embarrassing data to anyone. This exciting development introduces a new set of modeling possibilities using a richer data set but one that has smaller data coverage.

In general, secure data aggregation techniques in the cybersecurity sector produce aggregated data on security posture, control failures, incident frequencies, and losses. The available mathematical analytic tools include sums, averages, and high-level visibility into the overall data distribution of the variables. Individual inputs and more detailed data are not available as a feature of these techniques to protect the privacy and security of the underlying data. Given these new developments, there is a need for modified cyber risk frameworks that can ingest and use these smaller but richer data sets.

One of the most exciting developments is the ability to aggregate data on security posture and incidents at the sector level. Focusing on the industry level allows a group of similar firms facing similar threats to essentially pool information to understand and compare against the relevant peer group. From a modeling perspective, focusing on a peer group with common threats, similar incident frequencies, and comparable loss amounts opens new analytical possibilities for holding certain elements constant across the group and exploring the impact of control adoption and security posture on risk estimates.

In this paper we propose a modified cyber risk modeling framework that incorporates newly available securely aggregated data. We introduce a new top-line variable in a standard cyber risk model called the "Defense Gap Index" that measures how a firm's deviation from the average security posture, based on historical industry data of the peer group, impacts an organization's own security risk. We show further how to construct this gap measurement from the outputs of secure data aggregations done within a specific sector. Figure 1 introduces the proposed risk model that uses data collected from the sector to estimate the probability of a significant event in a given year (P), the average observed financial losses in the peer group (L), and now the gap index that relates control deviations from the group average to changes in risk outcomes.

**Figure 1: PLG = R**

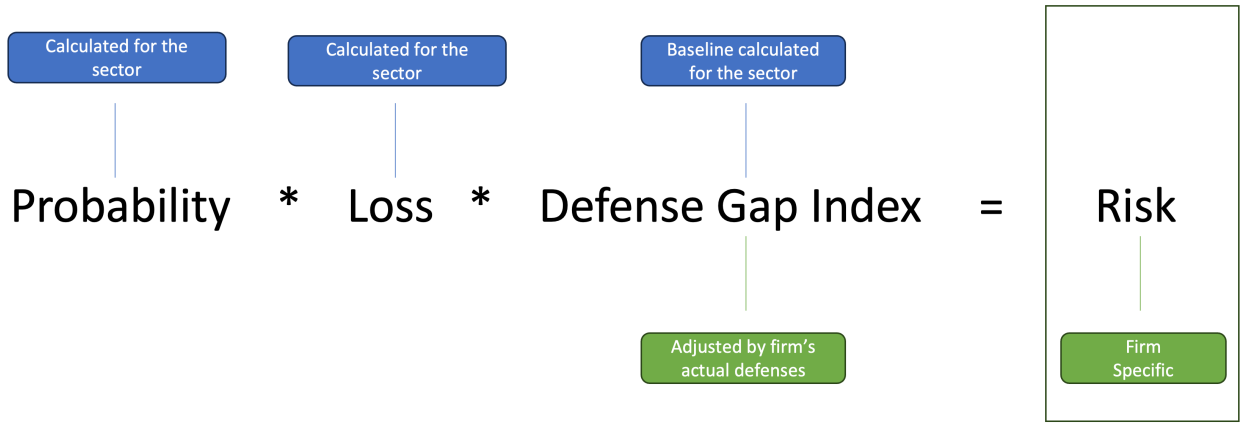

We apply this approach in a specific sector, in partnership with an industry ISAO[3], using actual data collected from 25 large firms with combined revenues of over USD 23 billion. The result is a general risk model for the industry and new private benchmarking tools for individual firms that allow them to answer the two outstanding questions – 1) what is my organization's estimated cyber risk and 2) how does it compare to the peer group?

## Related work
### Cyber risk modeling approaches
Some of the earliest work on what we now call cyber risk modeling was focused on the risk of data processing. In the 1970s, Courtney posited that risk to electronic data processing systems can be summarized with two elements – a statement of impact from a "difficulty" and the probability of encountering that difficulty (Courtney 1977). Nearly 50 years later, the basic formula for calculating risk is still widely used, although with different names.

The widespread adoption of computers throughout the business world in the 1990s and the growth of Internet connectivity later in the decade and throughout the 2000s highlighted the need for new information security protections. Markets responded and new risk transfer options in the form of cyber insurance appeared from companies such as Chubb, AIG, Lloyds, and Marsh (Gordon, Loeb, and Sohail 2003). In 2003, Gordon, Loeb, and Sohail published a framework for using insurance for cyber risk management that assesses risks, deploys security controls to mitigate some of the risk, and transfers remaining financial risk via insurance (Gordon, Loeb, and Sohail 2003). Researchers began exploring the decision-making process for businesses to transfer cyber risk via insurance (Mukhopadhyay et al. 2005). Around the same time, researchers began questioning whether the insurance market for cyber insurance was actually sustainable given the correlations among losses, the lack of actuarial data, and the difficulty of substantiating claims (Wang and Kim 2009b; 2009a; Böhme 2005; Baer and

---

[3] The data was collected from 25 large firms in the United States with combined annual revenues of USD 23 billion. Due to the sensitive nature of the results, we are keeping the name of the ISAO undisclosed in this version of the paper.

Parkinson 2007). The question of cyber insurability remains an important research area (Biener, Eling, and Wirfs 2015).

A new line of cyber risk research emerged in the early 2010s focusing on data breaches and the number of lost records to compare and quantify cyber incidents (Ayyagari 2012; Edwards, Hofmeyr, and Forrest 2016). Since the financial impact of data breaches were not available outside of insurance providers, researchers attempted to use the number of records exfiltrated as part of incident as a way to compare and quantify cyber breach (OECD 2013; Wheatley, Maillart, and Sornette 2016). The approach proved difficult because different records have different value and some of the quantification methods such as looking for changes in market capitalization tended to revert to the traditional growth path over time. Recent work is re-exploring the potential to estimate the value of data by estimating the value individuals put on access to their computer files (Cartwright, Cartwright, and Xue 2021).

Later in the 2010s, cyber risk modeling began splitting into two camps – those with access to large data sets such as insurance providers, and individual organizations that needed to understand and manage their own risk. Insurance providers such as brokers and underwriters arguably have access to the most detailed data on frequencies, and losses, but lack information on security posture within an organization.

Individual organizations have a much better understanding of their own security posture than their insurers do (asymmetric information), but they lack vital information about the broader cybersecurity landscape and information on incidents in their own sector that are valuable for forecasting their own risk. Since quantitative data is largely unavailable to individual firms, they rely heavily on heat maps and other qualitative measures to evaluate and address their cyber risk (Fink et al. 2009; Staheli et al. 2014; Jiang et al. 2022).

In the mid 2010s, two influential books appeared targeting individual organizations looking to quantify their own risk. Freund and Jones developed a bottom-up cyber risk modeling framework called Factor Analysis of Information Risk (FAIR) that has the same top-level structure as the model proposed by Courtney in 1977 (Courtney 1977; Freund and Jones 2014). The FAIR approach expands this into a taxonomy and ontology for building cyber risk models and quantifying cyber risk within a firm based on its own internal data and information that can be gleaned from other sources (Freund and Jones 2014). Around the same time, Hubbard and Seiersen published a popular book entitled, "How to Measure Anything in Cybersecurity Risk" (Hubbard and Seiersen 2016). Both approaches target risk analysts in individual firms and rely heavily on stochastic methods such as Monte Carlo simulations to estimate an organization's cyber risk.

At present, industry is moving toward risk quantification methods, and governments are making this a requirement in certain sectors, but the lack of external data sources remains a significant challenge.

## Cryptographic computing

Beginning in the 2010s and continuing into the 2020s, new privacy-enhancing technologies such as cryptographic computing began emerging that permit the collection, processing, analysis and sharing of information while protecting the confidentiality of the underlying data (OECD 2023). Advances in cryptography and expanding computational power unlocked the potential to do secure computations using homomorphic encryption that can compute functions over encrypted data (Abbe, Khandani, and Lo 2012; Asharov et al. 2012).

This has the potential to make new data sets available to researchers that were previously too sensitive to share into data aggregations. The technology is still developing but various use cases have emerged from double auctions in Denmark (Bogetoft et al. 2009), linking private data sets in Estonia (Bogdanov et al. 2016), protecting privacy in genome studies (Kamm et al. 2013), simulating electricity trading markets (Abidin et al. 2016) to estimating the gender wage gap using private wage data (Lapets et al. 2019). Current applications include privacy-preserving inventory matching systems for the banking sector (Polychroniadou et al. 2023) and distributed private attribution for advertising (Case et al. 2023).

In 2020, a cryptographic computing platform from MIT called SCRAM (Secure Cyber Risk Aggregation and Measurement) ran a secure multi-party computation to collect security posture, losses, and incident frequencies from six firms to produce new cyber security metrics that could be used for modeling in the future (de Castro et al. 2020). This was the first time that cryptographic computing was used to calculate previously unavailable cyber risk metrics. Now that the tools are available, the industry needs models that can use them.

In the cybersecurity context, we recommend encrypting data in transit and at rest, but assume that data must be decrypted during use. Cryptographic computing platforms are exciting because they bridge this final gap and allow the data to stay encrypted while in use.

## Data

Multi-party computation and encrypted data processing rely primarily on the ability to calculate sums over encrypted data (Abbe, Khandani, and Lo 2012). In the cyber risk context, sums are useful for counting the total number of incidents, arriving at a sum of total monetary losses, and counting the number of organizations that adopt a specific security control at a specific maturity level. The secure computation ingests values from a specific location (vector) within an encrypted spreadsheet that is contributed by a participating organization. Each of the encrypted elements is summed across the peer group and the resulting output (a new matching vector) is decrypted and contains the sum of each item in the input vector.

These sums can then be used to calculate averages for the group simply by dividing by the number of participants contributing data. Averages can be used for calculating the frequency of incidents and the average losses associated with an incident. Averages are typically calculated in post processing of the results data.

Another important data output from the computations are binary flags that are used for counting specific elements or creating distributions of variables across a set of data ranges. For example, binary flags are used to count the number of incidents that have a total monetary loss that falls between a specific range of values. These counts can then be combined to build a rough histogram of loss quartiles or quintiles that are then used to build the new gap index variable.

It is worth noting that individual records are not visible in computation results. Researchers cannot do traditional data cleaning on submitted data, but data checks are implemented with a verified checksum before data can be uploaded into the computation platform. The lack of visibility into individual inputs can lead to some imprecision in modeling the losses, for example, but this is the cost of increased privacy that is given to the input data.

**ISAO study:** For this paper, we securely collect data from 25 members of a single ISAO using the MIT SCRAM platform. The 25 organizations have combined annual revenue of over USD 23 billion. The collected data includes a rating of the maturity level of 22 controls in an organization, the number of incidents with losses larger than $5,000 between January 2021 and June 2023, information on which control failures are responsible for reported losses, and the total financial loss amount of security incidents during the relevant period. Specific details about the variables produced by the resulting computation are provided in the list below.

- **Maturity level (22 variables):** Average maturity level for each of 22 controls across the peer group (self-reported). Based on the Ransomware Readiness Index where all controls are drawn from the White House Executive Order on Improving the Nation's Cybersecurity, and the White House Memo to Corporate Executive and Business Leaders on Ransomware from 2021 (Spiewak, Reynolds, and Weitzner 2021).
- **Quartile flags – Maturity levels (88 variables)**: Count of maturity ratings for 22 controls over 4 potential responses (Not implemented, partially implemented, largely implemented, fully implemented). This provides a distribution of maturity levels across the participants.
- **Incident count (1 variable)**: Sum of the number of incidents across the peer group during the relevant period
- **Control failures (22 variables):** Count of the times individual controls failed leading to incidents with financial losses. Participants submitting an incident can implicate up to 5 failed controls as responsible for the reported financial loss.
- **Financial costs – total (1 variable):** Sum of the total financial costs across all incidents in USD.
- **Financial costs – by control (22 variables):** Data on the attributed costs of incident failures to for each of the 22 controls in USD. Data losses for a single reported incident are distributed evenly across all implicated controls in that incident.
- **Quintile flags – losses (5 variables):** Count of the number of incidents in each of five financial loss bands in USD. (1k-5k, 5k-50k, 50k-500k, 500k-5m, >5m)

These data are then used to build each of the components of the industry risk model and underpin the private tools that firms can use to compare their own security posture and risk to the peer group.

## Models and results

This modeling section details two modeling approaches that take advantage of the aggregated results for the sector. The first develops the PLG=R model and builds a new defense gap index (G) that captures the relationship between weighted security control deviations from the peer group and risk exposure. The second modeling section uses the same aggregated results to build an industry risk estimate and loss exceedance curve using a Monte Carlo simulation.

### Sectoral risk modeling approach 1: PLG = R

The PLG = R model can be re-written as follows to represent an organization's own risk relative to its peers.

**Equation 1** $\quad \bar{P}_{Peers} * \bar{L}_{Peers} * G_{Own} = AnnualRisk_{Own}$

**Equation 2** $\quad \bar{P}_{Peers} * \bar{L}_{Peers} * 1 = AnnualRisk_{Peers}$

Where:

**P** = Probability of an incident. Calculated as the average annual incident rate across the peer group. Once P is derived, it is held constant across the peer group under the assumption that similar firms face similar threats and defend similar assets.
**L** = Average financial loss amount per incident across the peer group. Once L is derived, it is also held constant over the peer group.
**G** = Defense Gap Index multiplier. The gap index represents how weighted security posture deviations from the peer average affect risk forecasts. The calculation of G is defined in detail in the following sections.
**Annual Risk** = The forecasted annual financial risk (expected value).

Equation 1 and Equation 2 above include a measure of frequency (P) of incidents and their impact (L) but introduce a new top-line element called the Defense Gap Index (G). The key innovation of the Defense Gap Index is that it uses actual loss data from the peer group, control failure attributions, and the average security posture of the peer group to estimate the relationship between weighted deviations from the average security control maturities of the peer group and changes in risk outcomes. This gives firms an empirically grounded means of predicting risk in the future to support investment decisions and can help enable regulators to set expectations for reasonable security posture.

In the well-known modeling approach Factor Analysis of Information Risk (FAIR), Freund and Jones capture the strength of security controls as "Resistance Strength" under the "Loss Frequency" category (Freund and Jones 2014). Control strength has an indirect effect on the model via the level of vulnerability the firm faces that impacts the frequency of successful attacks.

Since our core interest is understanding how changes in security posture affect cyber risk forecasts, our proposed model elevates differences in security posture from the peer group to a top line element in the risk model alongside probability and loss. The functionality of the Defense Gap Index is aligned with the goals of the variable "$Sec_T$" in Mukhopadhyay et al's CRAM model (Mukhopadhyay et al. 2019), but it is calculated differently and named as a "gap" index to capture the dynamic that higher scores of the variable relate to higher risk.

P and L are both derived from the secure computation as averages and represent the average probability of a significant incident for the peer group and the average monetary loss across all reported incidents.

In each step of the model explanation, we will use real-world data derived from the secure data collection done with 25 firms from a single ISAO. This allows us to illustrate the process while producing actual risk metrics and results for the sector.

ISAO data results from the secure computation:
- Average control maturity level: **78%** (high level, between largely and fully implemented)
- Number of incidents: **4**
- **P = 0.064** incidents per year per organization
- **L = $145,000** average loss per incident
- **G = 1** since this represents the average baseline weighted security of the peer group. In other words, the average security posture has no deviation from itself and is assigned a multiplier of 1.
- **R = $9,280** average annual cyber risk per firm (computed from PLG)
- Total losses: **$580,000**
- Implicated control failures: **5 controls implicated** across the total $580,000 of losses

*Defense Gap Index (G)*
At a high level, the Defense Gap Index acts as a multiplier that amplifies or reduces forecasted risk levels based on an organization's weighted deviations from the security control maturity averages of the peer group. The weights for specific controls are allocated based on actual financial losses attributed to control failures reported by members of the peer group.[4] Once individual control weights are assigned, the next step takes actual loss magnitudes contributed by the group and maps them to net weighted deviations from the group average. Large observed losses are assigned to large negative deviations (poorer security), while small observed losses are assigned to positive deviations from the average (better security). Next, we fit a function to the observed data points (including the known group average). This function is then used to calculate the Defense Gap Index multiplier (G). Using the gap index multiplier formula, organizations can privately input their own security posture to obtain a personalized Defense Gap Index multiplier (G) that goes into the PLG=R model to calculate their own risk.

---
[4] If an organization reports an incident, they must assign responsibility for the incident to specific control failures. They can implicate up to 5 control failures per incident. The reported loss amount is divided equally across all implicated controls.

Once the computation is complete and the Defense Gap Index calculation is parameterized, participants are sent the group values for P and L along with the Defense Gap Index formula. This allows them to privately do their own in-house risk modeling and answer the two outstanding questions of what is an organization's estimated risk exposure and how does its security compare with peers?

Figure 2 below provides a broad overview of the modeling approach where data from private computations in the first horizontal section feed into calculations of the Defense Gap Index in the second section. Finally, individual organizations can privately compute their own Defense Gap Index multiplier using their own security posture and use it for internal risk modeling. The five steps for modeling the Defense Gap Index multiplier are provided below and are populated with real-world data obtained from the computation with 25 members of an ISAO.

**Figure 2: Summary of the sectoral risk modeling approach**

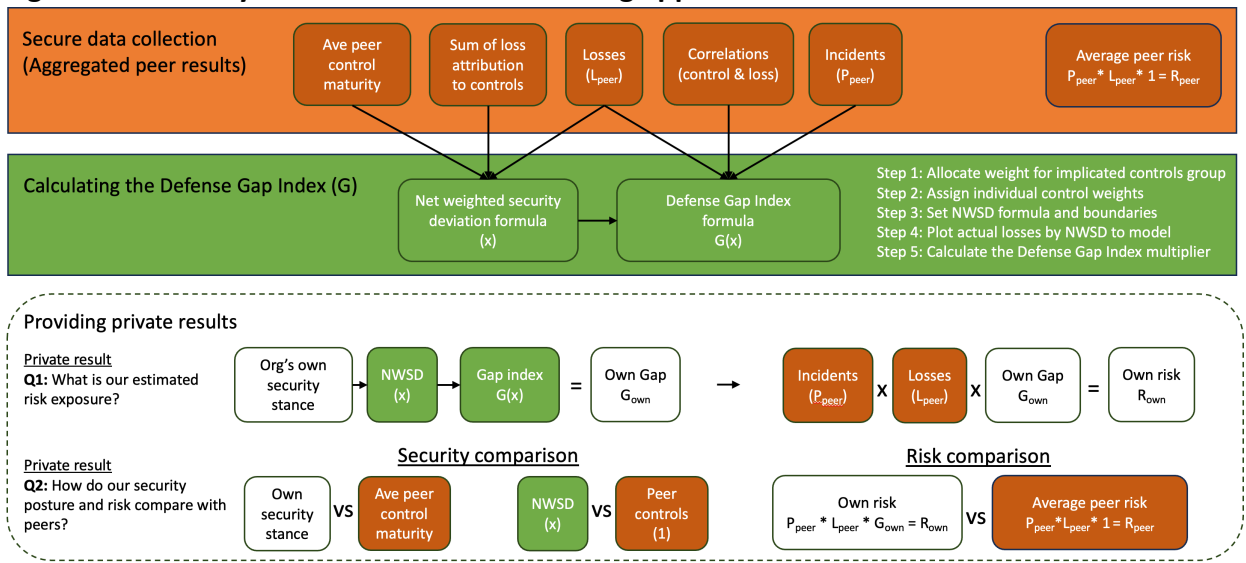

The next five steps explain how to derive the Defense Gap Index formula using data from the secure computation.

**Step 1: Allocate overall category weights between controls groups with and without losses**

In this first step, researchers building the industry model decide how much importance to place on control failures that lead to losses, the "loss group", compared to controls that are not implicated in loss events, the "no-loss group". Three possible weighting options are:

- Option 1: Equal weighting for all controls (unweighted)
- Option 2: Weighting based on the allocation of losses across implicated control failures

- Option 3: Weighting based on correlations of actual losses (or lack of losses) and security control maturity

Ideally the allocation of weights should be done using correlations between losses (and the lack of losses) and security control maturity while controlling for endogeneity. At present, the tools for allowing Option 3 are still under development, so this section describes how Option 2 is produced. This second option requires that some of the overall weight is assigned among controls implicated in failures, and the remainder is allocated across controls that have no associated losses. We start by considering an 85% (implicated) / 15% (non-implicated) split of the weights but then adjust to make sure that the smallest implicated weight is larger than any non-implicated weight.

ISAO result: We use a slightly modified data split of 75%/25% because of the wide loss range between the largest and smallest implicated control. We want to ensure the smallest implicated weight is larger than the non-implicated weights. Also, the relatively small number of implicated control failures (5) means additional weight should be added to the non-implicated controls.

**Step 2: Allocate individual control weights within loss and no-loss groups**
The second step allocates weights across individual controls in the loss and no-loss groups ( Figure 3). We assign controls in the "loss group" a high proportion of the total weight (e.g. 75%) and then the sub-weights of individual controls within the group are pro-rated based on the magnitude of losses assigned to each by the peer group. Sub-weights in the no-loss group are assigned as an equal distribution of the remaining weight (e.g. 25%).

ISAO result: There were 5 implicated controls with loss amounts in the ISAO data collection. The 5 implicated controls are assigned a combined 75% of the weight, while 17 non-implicated controls receive an equal share of the remaining 25%. The weights in the implicated group vary widely from 42% of the total weight assigned to "Evaluate employee skills" down to 1.9% of the weight assigned to "Deploy Multifactor Authentication". The full breakdown is available in (Table 2 in the Annex. These are all based on observed losses which ranged from largest amount of $325,000 attributed to employee skills to the smallest amount of $15,000 on MFA. Employee skills and training were the two largest loss areas followed by controls related to backup and then MFA. The remaining 17 controls each received an equal weight of 1.5%.
Figure 3 provides a breakdown of the ISAO weighting.

**Figure 3: ISAO weights applied to controls with and without attributed losses**

| **75%** of the total weight is assigned proportionately to the 5 controls with attributed losses based on the size of the observed financial impact.<br>Example: n = 5, weights vary from 42.0% to 1.9%.<br><br>**25%** of the total weight is assigned equally across all 17 non-implicated controls.<br>Example: n = 17, weights are all 1.5% | 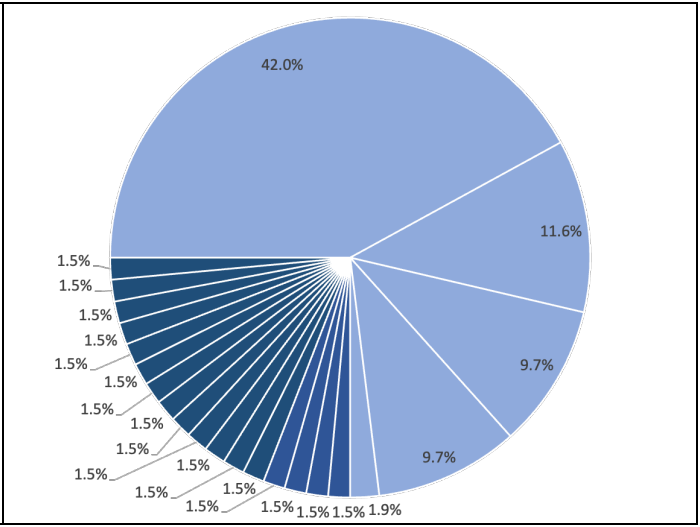 |
|---|---|

### Step 3: Net weighted security control deviation and boundaries

The third step uses the weights produced in Step 2 to create a net weighted deviation formula that individual organizations will use to calculate their own weighted security control deviations from the peer group. The group average and control weights are calculated in Step 2 above. The general equation is given in below:

**Equation 3**

$$NetWeightedDeviate = \sum_{n=1}^{i}((OwnMaturity_n/GroupAverage_n) * ControlWeight_n)$$

Where:
  OwnMaturity = The organization's own control maturity for control n
  GroupAverage = The average maturity level for control n across the peer group
  ControlWeight = The control weight assigned to control n in Step 2

In this step, the industry model developers determine a set of deviation boundaries that will be used to model the high and low ranges of observed losses. For example, the relevant ranges to consider could be 30% above and 30% below the group average, where having a net weighted security control deviation that is 30% below the average would correspond to the highest range of losses reported by the peers. The lowest range of losses are then assigned to net weighted security control deviations that are better than the peer average. Clearly there is some art involved in determining these ranges, but we have found +/- 30% to be a good set of modeling ranges in multiple sectors. In the future it should be possible to calculate the correlations between losses (and the lack of losses) and net weighted security control deviation to produce better estimates.

ISAO result: We select a maximum range of +/- 30% for the net weighted security deviation range to represent our best estimate of the weighted security variations across the sector.[5]

**Step 4: Create a model fitting observed loss data to the net weighted deviation for controls**
The fourth step evaluates and models the distribution of actual observed losses over the net weighted security control deviation boundaries defined in Step 3. We assume a non-linear, exponential model. We also assume that higher security (positive net weighted variation) corresponds to lower losses and vice versa (Eling and Wirfs 2019). The largest observed losses map to the lowest security levels (e.g. 30% below average) and the smallest losses to the higher security levels (e.g. up to 30% above average). The average loss and the average security level, which represent the averages of the peer group, correspond to a Defense Gap Index multiplier (G) of 1 and are used as one of the observations. Since individual losses are not visible, loss ranges in quartiles or quintiles provide the relevant data points for fitting the loss function. There is no precise way to place individual loss points from a range, but options include using the maximum, average, midpoint, or minimum as the representative point in the quartile. The average loss amount and average security level (corresponding to a Defense Gap Index of 1) are both known and serve as the grounding point for the model estimation.

ISAO result: The ISAO computation reveals four incidents reported by three firms spread over two quintiles. We calculate the loss model in this step by using the three observed loss amounts plus the computed average loss to build the loss model. The two incidents from the same firm are only visible to us as a single loss amount range which complicates interpreting the bands. The total losses across all incidents amount to $580,000. Two firms report losses between $50,000 and $500,000 and a third firm reports losses between $5,000 and $50,000.

For the higher loss quintile (50k-500k), we use $450,000 as the top end loss and assign it to a net weighted deviation of -30%. We arrive at the $450,000 number by subtracting away the bottom quartile's single highest loss ($50,000) from all reported losses ($580,000). We also know that the second firm's loss in the high quartile is larger than the upper limit of the smaller quartile ($50,000) so that can be subtracted as well – leaving us with $480,000. We used a slightly smaller $450,000 value to reflect the ambiguity around the actual loss amounts.

We know the average loss ($145,000) at the average level of security (net weighted security control deviation of 0). Finally, we assume the single loss in the lower quintile is close to the top of the range at $50,000 for an organization that has 15% better security than the average. These three available points provide us with enough data to estimate a curve that traverses through the average for the peer group in Figure 4.

---

[5] The maximum range is allowed to surpass 100% because weights on individual controls can vary considerably. In the ISAO case here, 42% of the total weight is assigned to one control (evaluating employee skills) so significant deviations of this single control can have large effects.

**Figure 4: Producing the Gap index scalar**

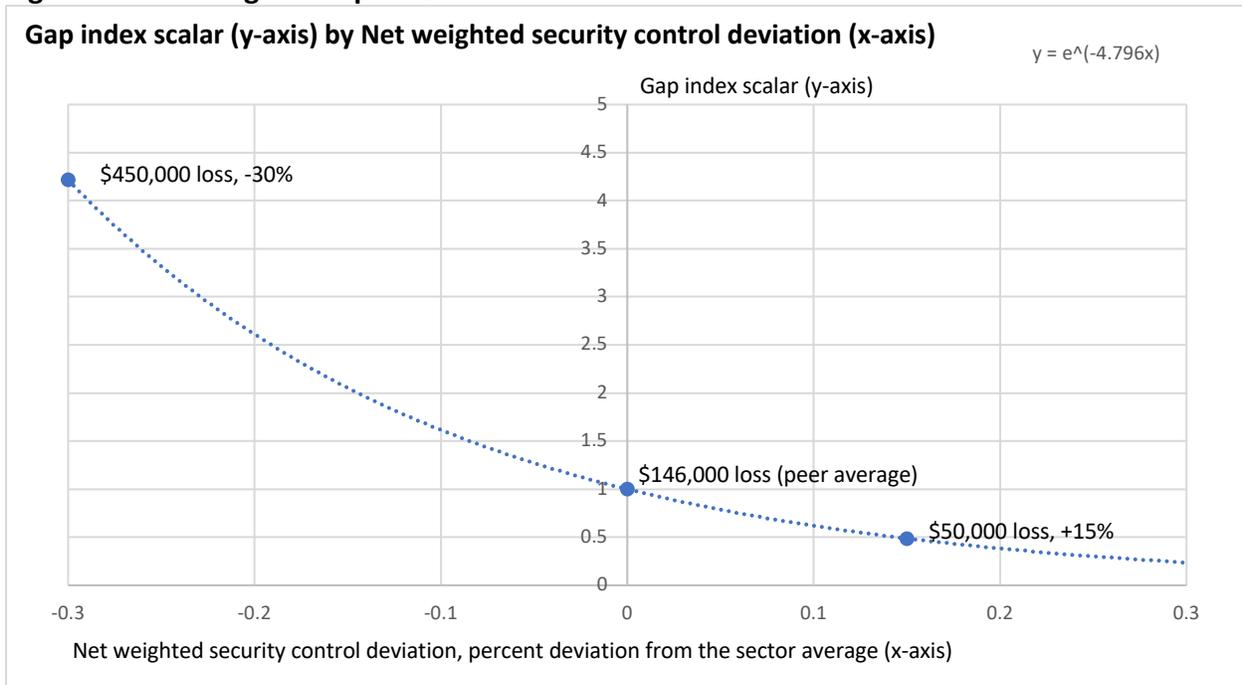

The resulting gap index model is:

**Equation 4** $$DefenseGapIndex = G = e^{-4.796*NetWeightedDeviate}$$

Where:
DefenseGapIndex = the Defense Gap Index multiplier (G) to be used in the equation P*L*G = R
NetWeightedDeviate = the net weighted security control deviation from the peer group's average control maturity.

*Computing private results*
Once the Defense Gap Index formula in Equation 4 is established and values for P and L are known from the computation results, organizations can now use the model to privately calculate their risk and compare their security posture and risk to their peers.[6] The equations and process for each are described in this section.

Q1: What is our organization's estimated risk exposure?
The process a participating organization uses to forecast their own risk exposure is seamless and automated via a spreadsheet once the secure computation results are available, but we step through the process in detail here.

---
[6] Participating organizations receive a results spreadsheet with detailed dashboards that only requires them to insert their own private values that were originally contributed into the secure computation that then populates all the dashboards.

The first step is calculating the organization's own risk exposure and comparing it to the peer group. This is done using Equation 3 to calculate its own net weighted security deviation (x). The results Equation 3 are then used in Equation 5 to calculate the organization's own Defense Gap Index value ($G_{Own}$}). Equation 5 should already include the derived constant value for the peer group that was calculated earlier for the entire group in Equation 4. Finally, the organization inserts $G_{Own}$ from Equation 5 into the two following risk equations, holding P and L constant, to obtain a forecast of its own annual cyber risk in monetary terms (Equation 6) and a forecasted incident size (Equation 7) in the case of an event. P and L are derived in the original secure computation and provided for the participants along with the DerivedConstant from the gap index modeling.

**Equation 3**
$$NetWeightedDeviate = \sum_{n=1}^{i}((OwnMaturity_n / GroupAverage_n) * ControlWeight_n)$$

**Equation 5** $\quad G_{Own} = GapIndexDefense = e^{DerivedConstant * NetWeightedDeviation}$

**Equation 6** $\quad AnnualRisk_{Own} = \bar{P}_{Peers} * \bar{L}_{Peers} * G_{Own}$

**Equation 7** $\quad ForecastedIncidentSize_{Own} = \bar{L}_{Peers} * G_{Own}$

### Q2a: How does our risk compare with our peers?

Once the organization knows its own forecasted annual risk and incident size, analysts can compare these results with the average results from the peer group. Equation 8 and Equation 9 compare the annual risk of the own firm with its peers, while Equation 10 and Equation 11 with the peer group on annual risk and forecasted incident sizes.

**Equation 8** $\quad AnnualRisk_{Own} = \bar{P}_{Peers} * \bar{L}_{Peers} * G_{Own}$

**Equation 9** $\quad AnnualRisk_{Peers} = \bar{P}_{Peers} * \bar{L}_{Peers}$

**Equation 10** $\quad ForecastedIncidentSize_{Own} = \bar{L}_{Peers} * G_{Own}$

**Equation 11** $\quad ForecastedIncidentSize_{Peers} = \bar{L}_{Peers}$

### Q2b: How does our security posture compare with our peers?

The next question that can be answered with the data is how the organization's own security posture compares with its peers.

There are two ways analysts can compare their organization's own security posture with peers in the sector (Figure 5). The first is using standard benchmarking tables outputs from the secure computation which provide the average maturity across the peer group for each control and a distribution of responses (not / partially / largely / fully implemented).

Analysts can also use the weighted controls lists that have been informed by actual losses across the group to accommodate control prioritization. The net weighted security control deviation measure provides a weighted comparison against the average (value of 1) of the peer group. For example, a net weighted deviation score of 0.75 implies that the organization's security posture is 25% lower than the sector's peer average after weighting each control by observed losses.

**Figure 5: Security posture comparison (unweighted and weighted controls)**

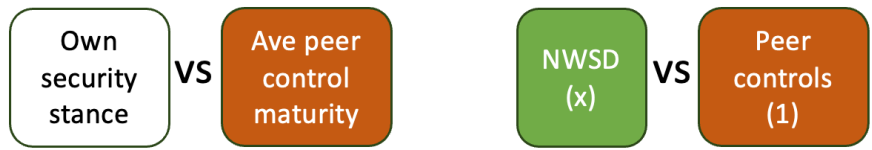

ISAO results for the industry

Equation 12    $AnnualRisk_{Peers} = \bar{P}_{Peers} * \bar{L}_{Peers} = 0.064 * \$145,000 = \$9,280$

Equation 13    $ForecastedIncidentSize_{Peers} = \bar{L}_{Peers} = \$145,000$

ISAO results for a particular firm

Equation 14    $G_{Own} = GapIndexDefense = e^{-4.796*NetWeightedDeviation}$

Equation 15    $AnnualRisk_{Own} = 0.064 * \$145,000 * G_{Own}$

Equation 16    $ForecastedIncidentSize_{Own} = \$145,000 * G_{Own}$

Using the ISAO results from the equations above, we illustrate how forecasted risk increases or decreases with changes in the net weighted security control deviation through the Defense Gap Index multiplier (G). Figure 6 shows annual expected risk based on variations in an ISAO member's defense posture. The average risk derived in Equation 12 of $9,280 per year for the average level of protection reflects the "fair price" for an insurance premium based on the incidents reported by the 25 firms over 2.5 years. However, if a member organization has substantially lower levels of control implementation, its forecasted annual loss could be over five times the average, or $49,723 as shown in Figure 6. At the other end of the control maturity spectrum, an organization with 35% higher weighted maturity will only suffer a forecasted average annual loss of $1,732, which is roughly one fifth lower than the average.

In insurance parlance, this fair price is the equivalent of the expected loss for the pool, but does not include other internal costs, external costs, economic profit needs, and capital costs that the insurance provider incurs to run its business. This means that the actual premium would need to be somewhat higher than the calculated expected loss for the insurance company to operate. The "fair price" calculation also assumes that all costs would be covered in the case of an incident, but that is typically not the case as there are exclusions and deductibles that lead to less than full coverage. The "fair price" calculations are imprecise, but they still provide a good starting point for organizations in the peer group to evaluate insurance offers.

**Figure 6: Annual cyber risk forecasts by net weighted security control deviation from group**

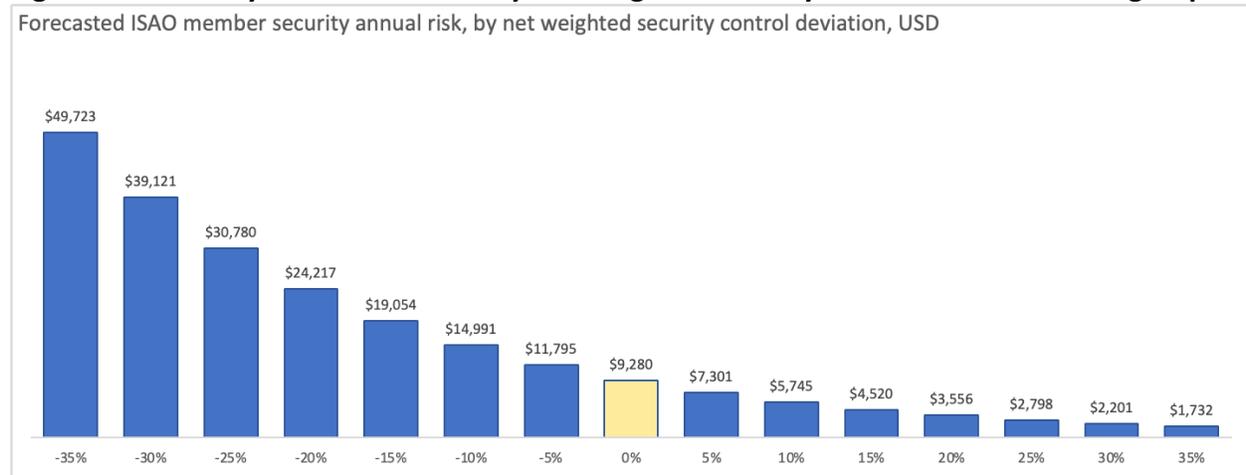

Figure 7 shows the same trend but forecasts the financial impact of an individual security incident based on the net weighted security deviation relative to the ISAO industry average. An organization with the average security posture could expect an incident size of $145,000 when there is a successful attack. However, organizations with a weighted net security gap that puts it 30% below average would expect an incident to cost $778,917 – nearly 5 times the average.

**Figure 7: Forecasted incident sizes by net weighted security control deviation from group**

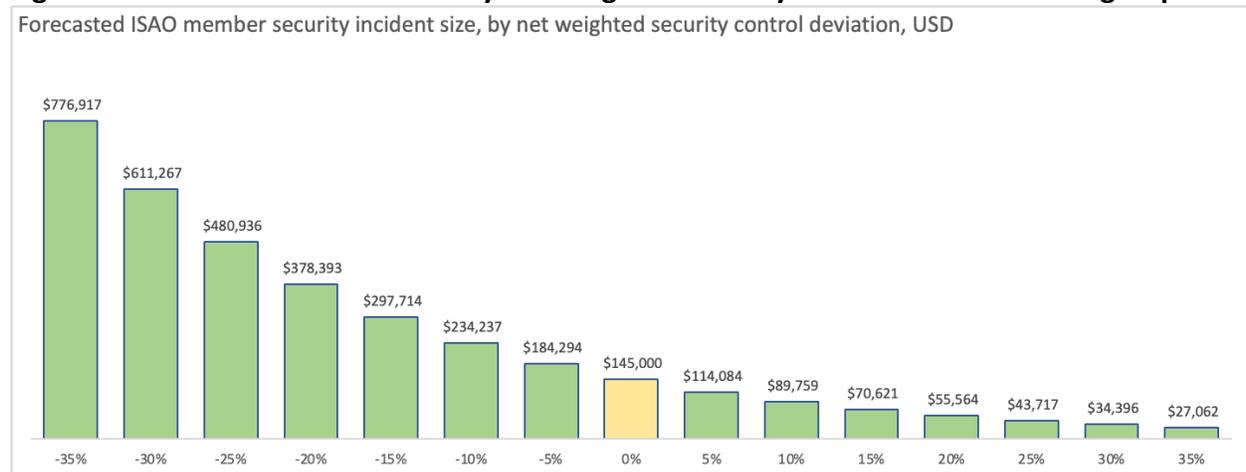

Sectoral risk modeling approach 2: Monte Carlo simulations & loss exceedance curves
*Monte Carlo simulations*

The peer data on losses can also be used in a Monte Carlo simulation at the peer group level to forecast the probability that the loss from a single cyber incident will be above a certain threshold. This requires an understanding of the distribution of financial losses across the group that can be gleaned from the secure computation loss quintiles. Eling and Wirfs use insurance data to study the costs of cyber events and find two categories of losses – the first they call the "cyber risks of daily life" with frequent but low financial losses, and the second that they call "extreme cyber risks" that are infrequent but have high associated losses (Eling and Wirfs 2019). One of their key findings is that the two categories of cyber events have different distributions and should be modeled separately.

Following this approach, we set up a Monte Carlo simulation based on the observed loss categories across the peer group. A mean, distribution, and probability are assigned to the large but infrequent loss category, and a different mean, distribution, and probability are assigned to the small but frequent loss category.

Our ISAO data show a potential cluster of one or two incidents in the $50,000 low end range and another potential cluster of incidents in the higher quintile in the $450,000 range. The computation results were ambiguous about the number of incidents in each quintile, so we will assume a 75% low end and 25% high end distribution that we have seen in other sectors. We model them separately within the same Monte Carlo simulation.[7] The Monte Carlo that selects losses distributed around $50,000 for 75% of the time and around $450,000 for the remaining 25% of the time (Table 1). We flatten the distributions by increasing the standard deviations for each of the categories to roughly correspond with the +/- 30% net weighted security deviation scores discussed earlier. The high distribution has a larger relative standard deviation indicating that losses at the high end will vary more than losses at the low end.

**Table 1: Monte Carlo inputs based on observed data**

| Variable | Low distribution | High distribution |
|---|---|---|
| Mean | $50,000 | $450,000 |
| Standard deviation | $25,000 | $300,000 |
| Probability | 75% | 25% |

The Monte Carlo simulation selects a random value 10,000 times that follows the distributions shown in Table 1 and represents the average security level for the group. For 75% of the time, that random value comes from the low distribution, and for 25% of the time from the high distribution. The results of the 10,000 iterations are then classified by their loss amount to provide a distribution of possible losses. In Figure 8, the Y axis shows the count of results in a specific range, and the X-axis shows the corresponding monetary loss.

---

[7] Spreadsheet equation: IF(RAND()<0.75,NORMINV(RAND(),50000,25000),NORMINV(RAND(),450000,300000))

The mean and spread of each distribution are determined by the observed data in the loss categories, but standard deviations are typically large and flat to cover the broad range of potential losses. The probability of a loss falling in either of the distributions should largely be set by the observed data but can be augmented with other known industry loss data if available. It is easier to introduce external data for this approach because no measure of the affected organization's security posture is required to place the loss in context.

Using the seeds from the peer group data collection, the next stop is running a Monte Carlo simulation with 10,000 or more instances. Random loss values cannot be negative, so any negative values are bottom censored at zero. This simulation represents expected losses based on the average security level for the peer group and is not tailored to a specific firm.

ISAO Results: The distribution emerging from Monte Carlo simulation using ISAO data is shown in Figure 8. clear peak is visible around $75,000 at the low end, while the distribution of high losses is thin and relatively flat around $500,000.

**Figure 8: ISAO Monte Carlo simulation of random loss values**

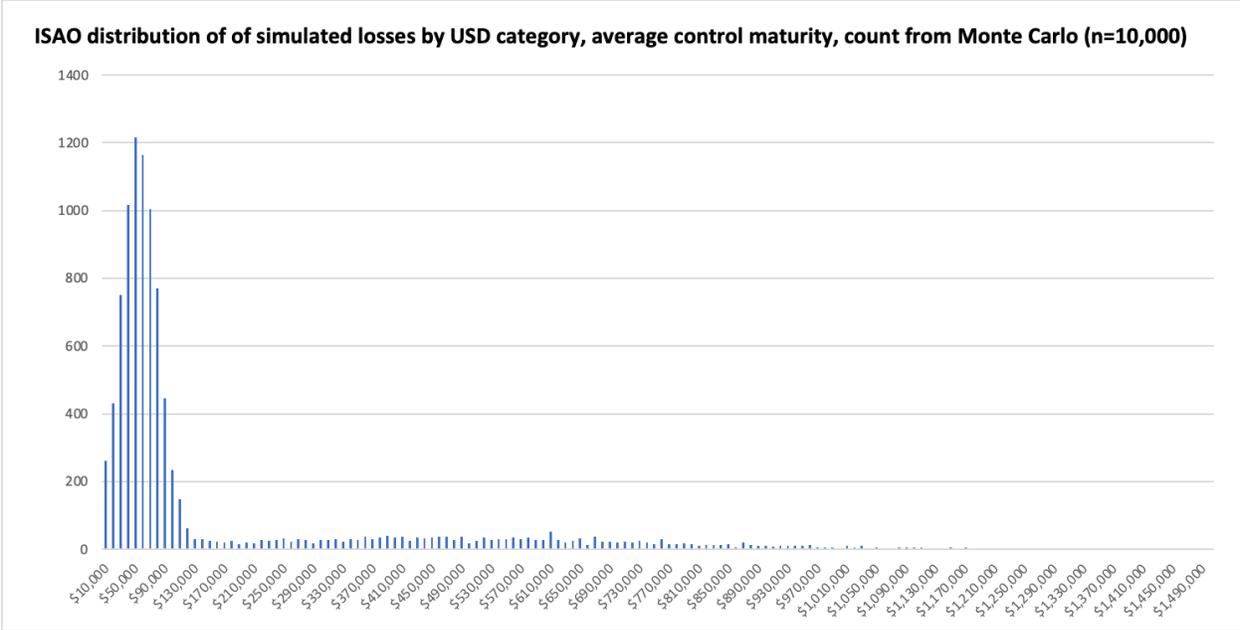

Note: Random values in the distribution cannot be negative and are bottom censored at zero.

*Loss exceedance curves*

The results of the Monte Carlo simulation can be used to build loss exceedance curves (also known as complementary cumulative distribution functions) that are commonly used in catastrophic risk modeling to describe the probability that a certain loss value will be exceeded in a predefined future time period (Grossi, Kunreuther, and Windeler 2005). Loss exceedance curves have also been adopted in cyber risk modeling to convey the probability that the losses from large cyber incident will exceed a given amount (Hubbard and Seiersen 2016), (Sokri 2019), (Humphreys 2021). They are useful for risk managers and governance boards charged with managing the organization's overall risk.

In the context of cyber risk governance, an organization's leadership may want to know whether the organization can handle the financial losses of a large incident and the probability that a single significant loss event will exceed a certain amount.

The loss data derived from secure computations is put into a Monte Carlo simulation whose outputs are use to create loss exceedance curves. We use a model based on (Hubbard and Seiersen 2016) and (Humphreys 2021) which shows the probability that a large incident will be above a certain loss threshold (Equation 17).

**Equation 17** $$LossExceedanceCurve = 1 - F_L(l) = p(L_{MC} > l)$$

Where:
$F_L$ is the cumulative distribution function of losses
$L_{MC}$ is the random variable of the loss from the Monte Carlo (real numbers).
l (lowercase) is the potential loss amount
In this implementation, the loss variable $L_{MC}$ represents the size of a single incident.

Figure 9 shows the ISAO peer group's loss exceedance curve based on the Monte Carlo simulation above. The results show that in 97% of cases, the cost of a significant incident will be over $10,000. The probabilities fall as the losses increase so that the probability of having a loss over $500,000 falls to 12%, and the probability of having a loss over $1,000,000 is only 1%. It is important to note that this simulation may only be representative for the peer group and not the entire sector due to selection bias issues.

**Figure 9: ISAO imputed loss exceedance curve**

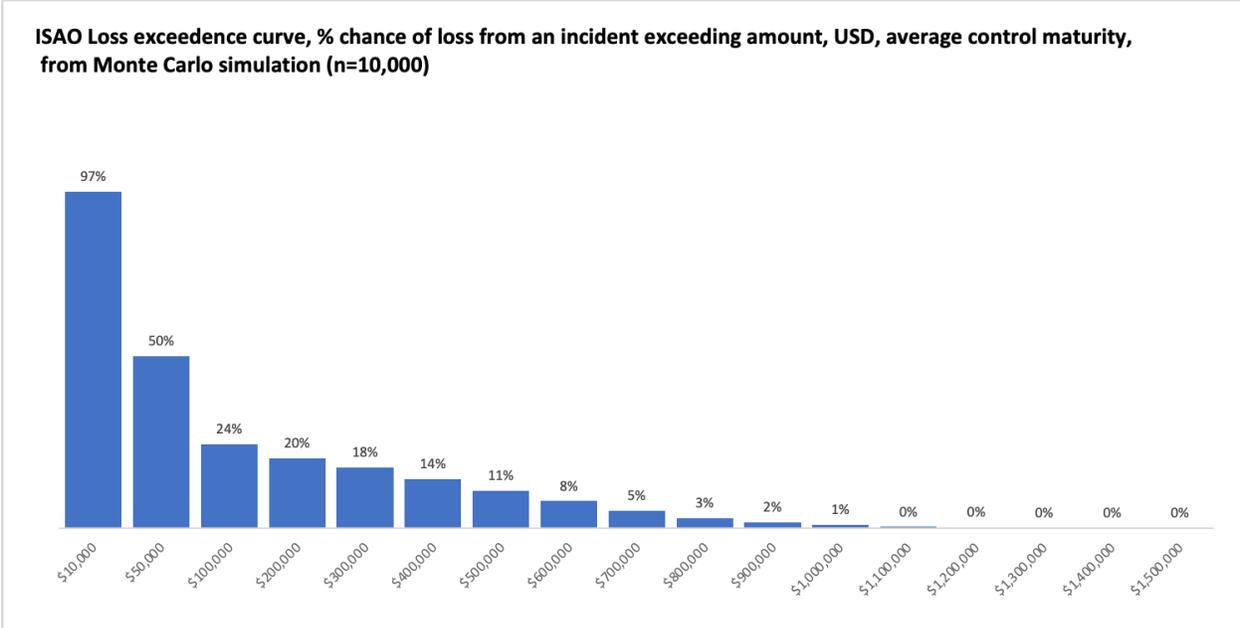

## Limitations of the work

These cyber risk models making use of secure computation results will improve our understand on risk and produce better estimates. Yet, there are several limitations to this research approach. First, the data that is used in the secure computations is self-reported by the organizations themselves. Although every effort is made to educate the participants about evaluating control maturities and estimating loss amounts, the self-reported data is likely to have variability that limits the precision of the results. In the future, automated data collections of specific variables could help minimize this challenge.

The risk modeling process, and the production of the Defense Gap Index in particular, require a bit of art mixed with science to locate and map reported loss ranges to net weighted deviations from the peer average. We understand that the process is imprecise, but we believe that perfection should not be the enemy of the good and having a small amount of actual data to model cyber risk for a peer group is better than having no data at all.

The number of organizations that can potentially participate in a computation is limited to a single sector, so there will be fewer incidents that are available for modeling then would be possible using firms from a variety of sectors. We also need a large representative sample from the sector to get results that reflect the state of the sector as a whole. Our strategy of limiting the research to one sector at a time allows us to hold P and L constant and evaluate changes in the Defense Gap Index (G) and how they affect risk. Broadening to the entire economy would certainly increase the number of incidents that could be used to model, but the assumption that P and L remain constant would be much more difficult to make.

The secure data collections likely suffer from some selection bias. Any organizations that is a member of an ISAO and is willing to invest time participating in a secure data collection for understanding cyber risk better is also likely to be among the most proactive in defending their data and networks. The 25 firms from the ISAO that participated in the study were somewhat surprised by the loss results. They expected much larger losses than were reported by the group, and several participants suggested that the issue may be due to selection bias. As a result, inferences related to the findings of the ISAO risk modeling should be limited to the profile of leading firms in the sector with regards to their security.

## Policy implications

The creation of the new defense gap index has important implications for policy making. First, it provides a valuable tool for organizations to calculate their cyber risk and compare it against their peer group in a way that has never been possible before. Second, it introduces a quantification methodology for prioritizing security controls based on actual losses and control failures reported by the peer group – providing clear guidance to policymakers on areas of particular need and targets for policy attention. Third, the defense gap index provides a holistic view of an organization's cyber security posture relative to its peers in the sector. Fourth, the gap index provides a baseline security posture for an industry that can be tracked over time to understand the sector's evolving security landscape.

This research shows that there are methods for calculating cyber risk metrics and models for specific sectors that can take advantage of new data coming from secure aggregations. These new secure computational techniques have opened a rich set of metrics that can be used to gauge the risk profile of a specific economic sector and allow organizations within that sector to compare themselves to their peers. Government efforts to bring together peer groups to jointly and securely aggregate cyber risk data could help policy makers and the organizations themselves obtain a much better understanding of cyber risk throughout the sector.

One of the key challenges in cyber risk modeling is a lack of standardized definitions and terms that are used across the industry. Until now, there has been limited effort to standardize the terminology since the data was previously too sensitive to share. But this is changing, and governments working with industry groups and academic researchers can play a role in helping standardize the definitions and terms we use for cyber risk modeling.

One of the key findings emerging from the ISAO data and backed up by other literature is that improving security for organizations that are significantly below the peer average can have an outsized effect relative to the investment. These firms with the lowest security levels offer the largest return on security investment because of the observed non-linearity of security losses.

Another related finding is that focusing interventions first on security control failures associated with the largest losses will likely have a larger return on investment and attention. Governments should prioritize research into uncovering better information about the effectiveness of controls to guide their own security investments and priorities.

## Conclusions & future work

The goal of this research is providing new models and data to answer two key questions that organizations have struggled to answer. What is an organization's estimated risk exposure? How does the security of an organization compare with its peers in the sector?

We provide the tools to answer each of these questions through the key innovation in this paper - a new variable called the Defense Gap Index in the top line of the risk model. The Gap Index works as a multiplier to increase or decrease forecasted risk for an individual firm based on the net weighted distance of its own security posture from the average security posture of the peer group. These comparisons are made possible using cryptographic computation tools.

This modeling approach provides new tools to individual organizations to forecast and benchmark their risk, but also allows policymakers to compare aggregate security levels across sectors. In the paper we apply the model to a data collection across 25 large firms in single sector to produce a benchmark for the industry and create powerful new tools for the participants to privately compute their own results.

Using data derived from a secure multi-party computation, we can develop a risk model for an ISAO sector and provide modeling tools to the participating firms to forecast their own risk based on their unique security posture, and then compare themselves to their peers. The model proposed in this paper is used for a secure data collection with an ISAO to build benchmarks of security posture, and risk models for the industry and individual firms.

Future research in this area should expand to additional sectors using similar methods so that the results could be compared to one another. Another area for future research would be developing new methods for introducing external data from outside the peer group into the modeling process for the Defense Gap Index.

Annex

**Table 2**: Observed losses and prorated control weights for the Gap Index (defense)

| Control | Observed losses | Equal control weights | Prorated control weights by losses: 75% prorated across losses 25% equally across non-losses |
|---|---|---|---|
| 5a. Eval employee skills | $325,000 | 4.5% | 42.0% |
| 5b. Deliver regular training | $90,000 | 4.5% | 11.6% |
| 6b. Test backups | $75,000 | 4.5% | 9.7% |
| 6d. Store backups offline | $75,000 | 4.5% | 9.7% |
| 1a. Deploy MFA | $15,000 | 4.5% | 1.9% |
| 2a. Deploy EDR | $0 | 4.5% | 1.5% |
| 2b. Hunt malicious activity | $0 | 4.5% | 1.5% |
| 3a. Encrypt in transit | $0 | 4.5% | 1.5% |
| 3b. Encrypt at rest | $0 | 4.5% | 1.5% |
| 4a. Remove sharing barriers | $0 | 4.5% | 1.5% |
| 4b. Threat intelligence | $0 | 4.5% | 1.5% |
| 6a. Regular backups | $0 | 4.5% | 1.5% |
| 6c. Protect backups | $0 | 4.5% | 1.5% |
| 7a. Timely updates & patching | $0 | 4.5% | 1.5% |
| 7b. Centralized patch system | $0 | 4.5% | 1.5% |
| 7c. Risk-based patching | $0 | 4.5% | 1.5% |
| 8a. Codify incident response plan | $0 | 4.5% | 1.5% |
| 8b. Test incident response plan | $0 | 4.5% | 1.5% |
| 8c. Maintain incident response plan | $0 | 4.5% | 1.5% |
| 9a. External pen testing | $0 | 4.5% | 1.5% |
| 9b. Red team exercises | $0 | 4.5% | 1.5% |
| 10a. Network segmentation | $0 | 4.5% | 1.5% |